\documentclass[%
 reprint,
superscriptaddress,
longbibliography,
color,
 amsmath,amssymb,
 aps, prd, 
]{revtex4-1}
\usepackage{graphicx}
\usepackage{subcaption}
\usepackage{caption}
\usepackage{float}
\usepackage{dcolumn}
\usepackage{bm}
\usepackage[colorlinks,linkcolor=blue,citecolor=blue,urlcolor=blue]{hyperref}

\begin{document}

\title{Constraining Lorentz and parity violations in gravity with multiband gravitational wave observations}

\author{Zhi-Xin Jia}
 \affiliation{School of Fundamental Physics and Mathematical Sciences, Hangzhou Institute for Advanced Study, UCAS, Hangzhou 310024, China.}
 \affiliation{Institute of Theoretical Physics, Chinese Academy of Science, Beijing 100190, China}
 \affiliation{University of Chinese Academy of Sciences, Beijing 100049, China}
 
\author{Tao Zhu}
\email{zhut05@zjut.edu.cn: Corresponding author}
\affiliation{Institute for Theoretical Physics and Cosmology, Zhejiang University of Technology, Hangzhou, 310032, China}
\affiliation{United Center for Gravitational wave Physics (UCGWP), Zhejiang University of Technology, Hangzhou, 310032, China}

\author{Zhoujian Cao}
\email{zjcao@bnu.edu.cn: Corresponding author}
\affiliation{School of Fundamental Physics and Mathematical Sciences, Hangzhou Institute for Advanced Study, UCAS, Hangzhou 310024, China.}%
\affiliation{Department of Astronomy, Beijing Normal University, Beijing 100875, China}%
\affiliation{Institute for Frontiers in Astronomy and Astrophysics, Beijing Normal University, Beijing 102206, China}

\date{\today}

\begin{abstract}

This study evaluates the capability of future multiband observations of gravitational waves emitted from binary black hole coalescences, utilizing joint third-generation ground-based (CE, ET) and space-based (LISA, Taiji, TianQin) detector networks, to constrain parity and Lorentz symmetry violations in the gravitational sector. We model these effects through a parameterized waveform framework that incorporates a set of parameters that quantify potential deviations from general relativity. The frequency-dependence of their effects is described by power-law indices $\beta$ (i.e., $\beta_{\bar \nu}$, $\beta_{\bar \mu}$, $\beta_{\nu}$, and $\beta_{\mu}$).
By analyzing events such as a high-signal noise ratio (SNR) ``golden event" like GW250114 and a massive binary system like GW231123 (total mass $190-265 M_\odot$) using two networks of ground- and space-based detectors, we demonstrate that multiband observations can significantly improve the current constraints on Lorentz and parity violations by several orders of magnitude, for both high-frequency ($\beta > 0$) and low-frequency ($\beta < 0$) modifications. The results reveal that while the exceptional SNR of the GW250114-like event yields superior constraints for high-frequency modifications ($\beta > 0$), the massive nature of GW231123 provides more stringent limits for low-frequency effects ($\beta < 0$). 
This work highlights the critical value of future multiband gravitational wave astronomy for conducting precise tests of general relativity across diverse binary populations.

\end{abstract}

\maketitle


\section{\label{sec: 1}Introduction}

The gravitational waves (GWs) signals from binary black hole mergers, directly detected by the LIGO-Virgo-KAGRA collaboration (LVK), have inaugurated a new era in gravitational physics and astonished astrophysicists with the existence of surprisingly massive black holes \cite{LIGOScientific:2016aoc, KAGRA:2021vkt, LIGOScientific:2020iuh, LIGOScientific:2016vbw, LIGOScientific:2016vlm}. The GW radiated by these events carries invaluable information about the local spacetime properties of compact binary systems. This enables us to test general relativity (GR) in the extremely strong and highly dynamical field regime \cite{KAGRA:2021duu}.

Building upon the remarkable success of ground-based detectors, space-based GW interferometers such as the Laser Interferometric Space Antenna (LISA)\cite{LISA:2017pwj}, Taiji \cite{Ruan:2018tsw, Hu:2017mde}, TianQin\cite{TianQin:2015yph, Liu:2020eko, Wang:2019ryf, Luo:2020bls} and DECi-hertz Interferometer Gravitational wave Observatory (DECIGO) \cite{Kawamura:2006up, Carson:2019kkh} are scheduled for launch in the mid-2030s. These detectors are designed to target GWs in the millihertz frequency band, complementing both the existing ground-based LVK network and next-generation ground-based detectors like Cosmic Explorer (CE) \cite{Evans:2021gyd} and Einstein Telescope (ET) \cite{Branchesi:2023mws}. This multiband approach presents an unprecedented opportunity to test foundational physics of GR across the GW spectrum \cite{LIGOScientific:2016aoc,2016PhRvL.117e1102V}.

In this context, typical stellar-mass or intermediate-mass binary black hole systems, first inspiral in the millihertz band for several years, where they are detectable by the space-based detectors, and subsequently reappear in the ground-based detector band several weeks before merger \cite{Gerosa:2019dbe}. During the prolonged observation in the millihertz band, space-based detectors enable highly precise measurements of the binary's component masses and sky location; conversely, in the ground-based detector band, the high signal-to-noise ratio (SNR) facilitates superior measurement of the GWs amplitude \cite{2016PhRvL.116w1102S}. Previous studies have highlighted the significant advantages of multiband GW astronomy. For instance, \cite{Grimm:2020ivq} showed that joint observations using space-borne detectors (such as LISA or B-DECIGO) and ground-based interferometers (such as ET) can effectively break parameter degeneracies. They found that while space-based detectors provide precise chirp mass estimates from the long-duration early inspiral, the addition of merger data from ground-based detectors vastly improves the estimation of individual masses and source positions, particularly for stellar-mass and intermediate-mass black hole binaries. In the context of multiband synergy, ref.~\cite{2016PhRvL.117e1102V} quantified how low-frequency observations from space-based detectors can inform and improve high-frequency ground-based analyses. It is found that using precise mass and localization priors from eLISA significantly breaks parameter correlations in ground-based data, leading to much tighter constraints on the primary black hole's spin and the remnant's properties, particularly for systems with asymmetric mass ratios. Highlighting the complementary nature of multiband astronomy, \cite{2019PhRvD.100f4024G} showed that a network of space-borne (LISA, B-DECIGO) and ground-based (LIGO/Virgo, ET) detectors can effectively probe deviations from GR across different dynamical regimes. Their study revealed that while space-based detectors are crucial for constraining low-frequency pre-Newtonian effects (such as dipole radiation in EdGB and Brans-Dicke theories) \cite{Yunes:2011we, Will:2014kxa}, ground-based detectors are essential for bounding high-frequency corrections (such as those in dCS gravity)\cite{Yagi:2011xp}, with the deci-hertz detector B-DECIGO playing a pivotal role in bridging these frequency bands to yield the tightest constraints \cite{Berti:2015itd}.

Parity symmetry and Lorentz symmetry are fundamental symmetries within GR. The potential violation of these symmetries arises in theoretical frameworks seeking to unify GR with quantum theory. Violation of parity symmetry has been experimentally discovered in the theory of weak interactions \cite{PhysRev.104.254, Wu:1957my}, while Lorentz symmetry is tested with exceptionally high precision under the Standard Model of particle physics \cite{Mattingly:2005re, Kostelecky:2008ts}. However, testing these symmetries within gravitational theories falls far short of achieving comparable levels of precision. Nevertheless, the quest for a unified theory of quantum gravity provides compelling theoretical motivations to probe these fundamental symmetries with increasing rigor. In the framework of String Theory \cite{Kostelecky:1988zi, Kostelecky:1991ak}, for instance, Lorentz symmetry may not hold exactly at the Planck scale; mechanisms such as the spontaneous breaking of symmetry in string field theory or interactions in non-commutative geometry suggest that minute deviations could persist and manifest at attainable energy scales. Similarly, loop quantum gravity (LQG) \cite{Gambini:1998it, Burgess:2002tb}, which postulates a discrete, granular structure of spacetime, predicts that the propagation speed of gravitational waves may exhibit frequency dependence, thereby violating Lorentz invariance through modified dispersion relations. Beyond the quantum gravity paradigms, various modified gravity theories constructed to address the limitations of GR—such as explaining cosmic acceleration—also naturally incorporate symmetry violations. A prominent example is dynamical Chern-Simons gravity \cite{Lue:1998mq, Jackiw:2003pm, Alexander:2009tp}, which explicitly introduces parity violation through a scalar field coupled to the Pontryagin density, leading to amplitude birefringence in GW propagation. Consequently, utilizing GW observations to constrain these theoretical predictions has become a central pursuit in fundamental physics, offering a unique window into the validity of symmetries in the strong-field regime \cite{Yunes:2016jcc}.

Different Lorentz and parity violation mechanisms may induce distinct effects in GW propagation, such as frequency-independent effects, amplitude and velocity birefringence effects, frequency-dependent damping, and nonlinear dispersion effects \cite{Zhu:2023rrx, Qiao:2019wsh}. It is crucial to construct a unified framework for model-independent tests of symmetry-breaking parameters directly with GW data to characterize these diverse effects. To this end, ref.~\cite{Zhu:2023rrx, Zhao:2019xmm} constructs a systematic parametrized framework for characterizing GWs propagation with Lorentz- and parity-violating effects \cite{Zhu:2023rrx, Zhao:2019xmm}, which has been used in analyzing GW data to obtain stringent constraints on Lorentz and parity violations \cite{Wang:2020cub, Wu:2021ndf, Gong:2021jgg, Wang:2021gqm, Haegel:2022ymk, Gong:2023ffb, Guo:2025bxz, Wang:2025fhw, Bian:2025ifp, Zhu:2022uoq}. For other related works, see refs.~\cite{Qiao:2022mln, Chen:2022wtz, Ng:2023jjt, Zhao:2022pun, Yamada:2020zvt, Okounkova:2021xjv, Wang:2021ctl, ONeal-Ault:2021uwu, Niu:2022yhr, Zhu:2022dfq, Callister:2023tws, Lagos:2024boe, Huang:2024lzu, Xiong:2024vsd, Ghosh:2023xes, Nair:2024xdb} and references therein. Furthermore, other parametrization frameworks are discussed in \cite{Saltas:2014dha, Nishizawa:2017nef, Tahura:2018zuq, Ezquiaga:2021ler, Jenks:2023pmk}. Forecasts for constraining parity and Lorentz violation with future ground-based and space-based GW detectors have also been considered, see refs.~\cite{Zhang:2024rel, Lin:2024pkr, Zhang:2025kcw, Califano:2023aji, Hou:2024xbv, Mirshekari:2011yq, Liang:2023pbj, Hu:2020rub} as examples.

The main purpose of this paper is to investigate the capability of future multiband GW observation, combining third-generation ground-based detectors (CE, ET) and space-based detectors (LISA, Taiji, TianQin), to constrain parity and Lorentz violations in gravity. We employ a parametrized modified GW waveforms incorporating parity- and Lorentz-violating effects and conduct a series of Fisher information matrix (FIM) analyses for two representative binary black hole coalescences. Specifically, this paper is organized as follows. In Sec. \ref{sec:2}, we briefly introduce a unified parametrization of the modified GW propagation equation \cite{Zhu:2023rrx}. Within this framework, we also present the explicit dependence of the waveform on both intrinsic source parameters and the symmetry-breaking coefficients that are the targets of our constraints. Sec. \ref{sec: 3} outlines the methodology of multiband observations, detailing the detector networks and their complementary frequency coverage. The statistical framework based on the FIM \cite{Cutler:1994ys} is then presented; we use it to estimate the 90\% credible uncertainties on the symmetry-breaking energy scales \(M_{\mathrm{LV}}\) and \(M_{\mathrm{PV}}\). Our results are discussed in Sec. \ref{sec: 4}, and conclusions together with an outlook are given in Sec. \ref{sec: 5}.

\section{\label{sec:2}Modified waveforms with Parity- and Lorentz-violating Effects}

In this section, we introduce a universal parametrization framework designed to systematically characterize deviations from GR during the propagation of GWs \cite{Zhu:2023rrx, Zhao:2019xmm}. In this framework, the parametrized equation of motion for circular polarization modes of GWs is given by,
\begin{align}\label{eom_A}
h''_A + (2+\bar \nu + \nu_A) \mathcal{H} h'_A + (1+\bar \mu+ \mu_A) k^2 h_A=0,
\end{align}
where $A = {\rm R}$ and ${\rm L}$ denote the right- and left-handed circular polarization modes of GWs, respectively. In this expression, a prime indicates differentiation with respect to conformal time $\tau$, $\mathcal{H} = a'/a$ represents the conformal Hubble parameter (with $a$ as the cosmic scale factor), and $k$ is the wavenumber of GWs. Within this formalism, deviations from GR are fully encapsulated by four parameters: $\bar\nu$, $\bar\mu$, $\nu_A$, and $\mu_A$, each corresponding to specific physical modifications to GW propagations. These modifications are categorized into three classes: (i) frequency-independent effects ($\bar\nu$ and $\bar\mu$) which alter the propagation speed and friction; (ii) parity-violating effects ($\nu_A$ and $\mu_A$) which induce amplitude and velocity birefringences; and (iii) frequency-dependent effects ($\bar\nu$ and $\bar\mu$) resulting in non-linear dispersion and frequency-dependent damping rate. While the first category does not explicitly relate to symmetry breaking, the latter two represent the signatures of parity and Lorentz violations in gravity, respectively. A summary of these parameters in many modified gravity theories is provided in ref.~\cite{Zhu:2023rrx}. In this study, we focus exclusively on the parity- and Lorentz-violating effects.

The parameters $\nu_A$ and $\mu_A$ characterize the effects induced by the gravitational parity violation. Specifically, $\mu_A$ induces velocity birefringence, causing the two circular polarizations to propagate at different velocities and thus arrive at different times. And $\nu_A$ results in amplitude birefringence, which manifests as different damping rates for the left- and right-handed modes, leading to an increase or decrease in their relative amplitudes during propagation. For a broad class of parity-violating theories, these parameters are frequency-dependent and can be parametrized as:
\begin{align}
\mathcal{H} \nu_{\mathrm{A}} &=\left[\rho_{\mathrm{A}} \alpha_{\nu}(\tau)\left(k / a M_{\mathrm{PV}}\right)^{\beta_{\nu}}\right]^{\prime}, \\
\mu_{\mathrm{A}}&=\rho_{\mathrm{A}} \alpha_{\mu}(\tau)\left(k / a M_{\mathrm{PV}}\right)^{\beta_{\mu}},
\end{align}
where $M_{\rm PV}$ is the energy scale of parity violation, $\beta_\nu$ and $\beta_\mu$ are two power-law indices, and $\alpha_\nu, \alpha_\mu$ are arbitrary time-dependent functions. Here $\rho_{\rm R}=1$ and $\rho_{\rm L}=-1$.

Similarly, Lorentz symmetry violations lead to non-zero, frequency-dependent $\bar \nu$ and $\bar \mu$. Here, $\bar \nu$ introduces frequency-dependent friction, while $\bar \mu$ modifies the standard linear dispersion relation into a non-linear one. These effects are parametrized as:
\begin{align}
\mathcal{H} \bar{\nu} &=\left[\alpha_{\bar{\nu}}(\tau)\left(k / a M_{\mathrm{LV}}\right)^{\beta_{\bar{\nu}}}\right]', \\
\bar{\mu}&=\alpha_{\bar{\mu}}(\tau)\left(k / a M_{\mathrm{LV}}\right)^{\beta_{\bar{\mu}}},
\end{align}
where $M_{\rm LV}$ represents the Lorentz violation energy scale, and $\beta_{\bar \nu}, \beta_{\bar \mu}$ are the corresponding power-law indices.

By solving the modified equation of motion Eq.~(\ref{eom_A}), one can derive the explicit GW waveforms. The resulting amplitude and phase corrections to the standard GR waveform are expressed as:
\begin{equation}
\tilde h_A(f) = \tilde h_A^{\rm GR}(f) e^{ \rho_A \delta h_1 +\delta h_2} e^{i  (\rho_A \delta \Psi_1 + \delta \Psi_2)}.\label{waveforms}
\end{equation}
Here, $\tilde h_A^{\rm GR}$ is the reference GR waveform. The amplitude corrections $\delta h_1 = A_{\nu} (\pi f)^{\beta_\nu}$ and $\delta h_2 = -A_{\bar \nu} (\pi f)^{\beta_\nu}$ are driven by $\nu_A$ and $\bar{\nu}$, while the phase corrections $\delta \Psi_1$ and $\delta \Psi_2$ arise from $\mu_A$ and $\bar{\mu}$. Their analytical forms are given by:
\begin{align}
\delta h_1 &= A_{\nu} (\pi f)^{\beta_\nu}, \\
\delta \Psi_1 &= 
\begin{cases}
A_{\mu} (\pi f)^{\beta_\mu+1}, & \beta_\mu \neq -1, \\
A_{\mu} \ln u, & \beta_\mu =- 1,
\end{cases}
\end{align}
for parity-violating effects, and:
\begin{align}
\delta h_2 &= -A_{\bar \nu} (\pi f)^{\beta_\nu}, \\
\delta \Psi_2 &=
\begin{cases}
A_{\bar \mu} (\pi f)^{\beta_{\bar \mu}+1}, & \beta_{\bar \mu} \neq -1, \\
A_{\bar \mu} \ln u, & \beta_{\bar \mu} =- 1,
\end{cases}
\end{align}
for Lorentz-violating effects. The coefficients $A_{\nu}$, $A_{\bar \nu}$, $A_{\mu}$, and $A_{\bar \mu}$ are defined as:
\begin{align}
A_{\nu} &= \frac{1}{2} \left(\frac{2}{M_{\rm PV}}\right)^{\beta_\nu}\left[\alpha_\nu(\tau_0) - \alpha_\nu(\tau_e) (1+z)^{\beta_\nu}\right], \label{Anu}\\
A_{\mu} &= \frac{(2/M_{\rm PV})^{\beta_\mu}}{\Theta(\beta_\mu+1)}  \int_{t_e}^{t_0} \frac{\alpha_\mu}{a^{\beta_\mu+1}}dt, \label{Amu}\\
A_{\bar \nu} &= \frac{1}{2} \left(\frac{2}{M_{\rm LV}}\right)^{\beta_{\bar \nu}}\left[\alpha_{\bar \nu}(\tau_0) - \alpha_{\bar \nu}(\tau_e) (1+z)^{\beta_{\bar \nu}}\right], \label{Abnu}\\
A_{\bar \mu} &=  \frac{(2/M_{\rm LV})^{\beta_{\bar \mu}}}{\Theta(\beta_{\bar \mu}+1)} \int_{t_e}^{t_0} \frac{\alpha_{\bar \mu}}{a^{\beta_{\bar \mu}+1}}dt. \label{Abmu}
\end{align}
In these expressions, $t_e$ and $t_0$ denote the emission and arrival times of GWs at the source and detectors, respectively; $z$ is the redshift of the source; $f$ is the observer-frame frequency; and $u=\pi {\cal M} f$, where ${\cal M}$ is the measured chirp mass. The function $\Theta(1+x)$ equals $1+x$ for $x\neq -1$ and $1$ for $x=-1$. Given our focus on local GW events, we treat $\alpha_\nu, \alpha_{\mu}, \alpha_{\bar{\nu}}, \alpha_{\bar{\mu}}$ as constants. We adopt the following cosmological parameters: $\Omega_m=0.315$, $\Omega_{\Lambda}=0.685$, and $H_0=67.4\; {\rm km}\;{\rm s}^{-1}\; {\rm Mpc}^{-1}$.

\section{\label{sec: 3} Analysis with the modified Waveforms}

\subsection{Networks for Multiband Observations}

Multiband GW observations provide a critical opportunity to test GR and to constrain potential deviations from it across different dynamical regimes. Current ground-based detectors (e.g., LIGO/Virgo) observe the high-frequency inspiral and merger of binary black holes. Future space-based missions (e.g., LISA, Taiji, Tianqin) will detect the same systems years earlier, during their long, low-frequency inspiral. Next-generation ground observatories (ET, CE) will further extend the covered frequency range. This multiband coverage allows a single source to be tracked across its evolution, dramatically improving the precision of waveform measurements. Such multiband detection is especially powerful for probing Lorentz- and parity-violating effects, which may accumulate over time or imprint differently across frequency bands, offering complementary tests that are inaccessible to single-band observations alone.

In this paper, we study the constraints on Lorentz and parity violations in gravity by multiband observations with two multiband GW networks: (1) Network A consists of two third-generation ground-based detectors (ET and CE) and two space-based detectors (LISA and Taiji); (2) Network B consists of two third-generation ground-based detectors (ET and CE) and two space-based detectors (LISA and Tianqin). 

ET features a triangular design with a 10-km arm length to be established in Europe \cite{Branchesi:2023mws}. Complementing this, the CE is envisioned as a dual-site network: a 40-km observatory in the United States and a 20-km facility in Australia \cite{Evans:2021gyd}. LISA is a future space-borne GW detector led by the European Space Agency (ESA). It consists of three spacecraft forming an equilateral triangle with an arm length of 2.5 million kilometers. The constellation orbits the Sun in an Earth-trailing orbit (approximately $20^{\circ}$ behind Earth) \cite{LISA:2017pwj}. Taiji is a space-based GW mission proposed by the Chinese Academy of Sciences (CAS). Similar to LISA, it adopts a heliocentric orbit and consists of three satellites forming a triangle. However, Taiji is planned to have a slightly longer arm length (3 million kilometers) compared to LISA \cite{Ruan:2018tsw}. Tianqin is a Chinese space-based GW observatory project led by Sun Yat-sen University. Unlike the heliocentric LISA and Taiji, Tianqin utilizes a geocentric orbit (orbiting the Earth) with a constellation of three satellites. Its arm length is shorter, approximately 170,000 kilometers \cite{TianQin:2015yph}. We summarize the information of the five considered GW detectors in Table \ref{tab:detector_config}.

\begin{figure*}
\centering
\includegraphics[width=0.9\linewidth]{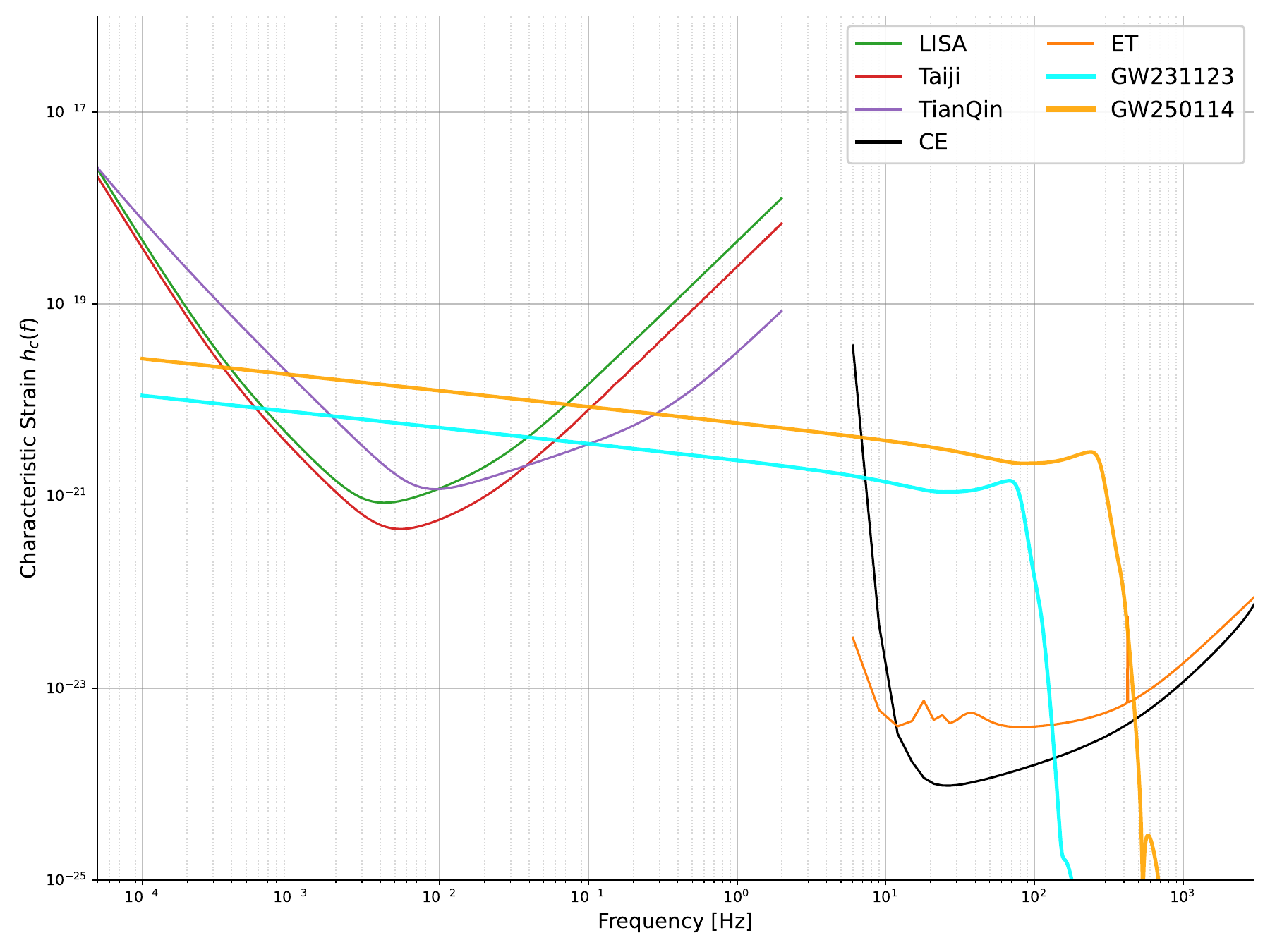}
\captionsetup{justification=raggedright,singlelinecheck=false}
\caption{\label{fig:psd} Sensitivity curves of various GW interferometers, plotted alongside the characteristic amplitude of GW250114 and GW231123. The figure illustrates the multiband nature of the event: the early inspiral phase is observable by space-based detectors, whereas the late inspiral, merger, and ringdown phases fall within the detection band of ground-based instruments.}
\label{five_detectors}
\end{figure*}

In Fig.~\ref{five_detectors}, we illustrate the noise power spectral densities (PSDs) of both ground-based and space-based detectors. The PSD data of two ground-based detectors, ET and CE, are from the official data files \cite{Hild:2010id, Reitze:2019iox}. The other three PSD data for the space-based detectors, Taiji, LISA, and Tianqin, are from the theoretical formula introduced below. The noise performance of the space-based GW observatories is primarily characterized by their instrumental sensitivity curves, which incorporate displacement measurement noise and residual acceleration noise. For the Taiji program, the noise PSD is modeled following ref.~\cite{Luo:2019zal, Liu:2023qap} as:
\begin{align}
S_{\rm n}(f) = &\frac{10}{3 L^{2}} \left(P_{\rm dp} + 2 \left(1+\mathrm{cos}^{2}\left(f/f_{*}\right)\right) \frac{P_{\rm acc}}{(2 \pi f)^{4}}\right) \nonumber \\
&\times \left(1+0.6 (f/f_{*})^{2}\right),
\end{align}
where $L = 3 \times 10^{9}$ m represents the interferometric arm length and $f_{*} = c/(2 \pi L)$ is the characteristic transfer frequency (in natural units). The instrumental noise budget is dominated by the displacement noise $P_{\rm dp}$ and the acceleration noise $P_{\rm acc}$, defined respectively as:
\begin{equation}
P_{\mathrm{dp}}=\left(8 \times 10^{-12} \mathrm{~m}\right)^2\left(1+\left(\frac{2\,\mathrm{mHz}}{f}\right)^4\right) \mathrm{Hz}^{-1},
\end{equation}
and
\begin{align}
P_{\mathrm{acc}} = &\left(3 \times 10^{-15} \mathrm{~m} \mathrm{~s}^{-2}\right)^2\left(1+\left(\frac{0.4\, \mathrm{mHz}}{f}\right)^2\right) \nonumber \\
&\times\left(1+\left(\frac{f}{8\, \mathrm{mHz}}\right)^4\right) \mathrm{Hz}^{-1}.
\end{align}
Similarly, the LISA configuration adopts a baseline of $L = 2.5 \times 10^{9}$ m. While sharing a comparable functional form for its sensitivity curve, LISA's displacement noise PSD, $P_{\mathrm{dpL}}$, is specifically characterized by a higher noise floor~\cite{Robson:2018ifk}:
\begin{equation}
P_{\mathrm{dpL}}=\left(15 \times 10^{-12} \mathrm{~m}\right)^2\left(1+\left(\frac{2\, \mathrm{mHz}}{f}\right)^4\right) \mathrm{Hz}^{-1}.
\end{equation}

In contrast, the TianQin observatory, operating in a geocentric orbit with a shorter arm length of $L = \sqrt{3} \times 10^{8}$ m, follows a distinct sensitivity model~\cite{TianQin:2020hid}:
\begin{align}
S_{\rm n}(f) = &\frac{10}{3 L^2}\left[S_{\rm x}+\frac{4 S_{\rm a}}{(2 \pi f)^4}\left(1+\frac{10^{-4} \mathrm{~Hz}}{f}\right)\right] \nonumber \\
&\times \left[1+0.6 \,(f/f_{*})^2\right].
\end{align}
For TianQin, the transfer frequency is defined as $f_{*} = c/(2 \pi L)$. The corresponding displacement measurement noise $S^{1/2}_{\rm x}$ and residual acceleration noise $S^{1/2}_{\rm a}$ are set to $1\times10^{-12}\ \mathrm{m} /\mathrm{Hz}^{1/2}$ and $1\times10^{-15}\ \mathrm{m} \ \mathrm{s}^{-2}/\mathrm{Hz}^{1/2}$, respectively. These distinct noise floors and frequency responses across Taiji, LISA, and TianQin provide a diverse set of baselines for probing symmetry-violating effects in the millihertz band.

\subsection{\label{subsec: 3.1} Simulated signals with modified waveforms}

We perform signal injections into simulated data for our detector networks using the \texttt{PyCBC} library \cite{PyCBC}. Our analysis is built upon two complementary GW events: a high SNR ``golden" binary black hole merger like GW250114 \cite{LIGOScientific:2025rid} and an exceptionally massive system like GW231123 \cite{LIGOScientific:2025rsn}. This selection is deliberate: the high SNR ($\approx 80$) of the GW250114-like event enables a precise test of GR through detailed waveform matching. In contrast, the GW231123-like event, with a total mass of $\sim 236, M_\odot$, has a primary component residing in the pair-instability mass gap, offering a unique laboratory to study intermediate-mass black hole formation and to probe waveform systematics in an extreme mass-ratio regime.

For parameter estimation, we adopt a modified waveform model, as described in Eq.~(\ref{waveforms}). The GR-based waveform $\tilde{h}^{\rm GR}_A$ is given by the \texttt{IMRPhenomXPHM} phenomenological approximant \cite{Pratten:2020ceb}, which models the inspiral-merger-ringdown of quasi-circular, precessing binary black holes in the frequency domain. To this GR template, we explicitly add phase and amplitude modifications that encode possible deviations induced by Lorentz and parity violations. We choose \texttt{IMRPhenomXPHM} as the underlying model because it incorporates higher-order multipole moments, $(2,1)$, $(3,3)$, $(3,2)$, and $(4,4)$, via a ``twisting up" of the non-precessing \texttt{IMRPhenomXHM} baseline \cite{Garcia-Quiros:2020qpx, Hannam:2013oca, Khan:2015jqa}. These higher harmonics help break degeneracies among source parameters \cite{CalderonBustillo:2015lrt}. Furthermore, the model employs efficient multibanding interpolation, making it computationally faster than time-domain effective-one-body counterparts while retaining accuracy against numerical-relativity simulations. 

\begin{table}
    \centering
    \caption{Parameters of the gravitational wave events.}
    \label{tab:gw_parameters}
    \begin{ruledtabular}
    \begin{tabular}{lcc}
        \textbf{Parameter} & \textbf{GW250114} & \textbf{GW231123} \\
        \hline
        Primary Mass $m_1$ ($M_\odot$)       & $33.7$  & $149$  \\
        Secondary Mass $m_2$ ($M_\odot$)     & $32.3$  & $90$   \\
        Chirp Mass $\mathcal{M}$ ($M_\odot$) & $28.7$  & $100$  \\
        Effective Spin $\chi_{\text{eff}}$   & $-0.03$ & $0.01$ 
    \end{tabular}
    \end{ruledtabular}
\end{table}

The sampled parameter space is
\begin{equation}
\label{eqn:parameters}
\theta = \{ \mathcal{M}_c, \chi_{\rm eff}, t_0, \phi_0, C\},
\end{equation}
where $\mathcal{M}_c = (m_1 m_2)^{3/5} / (m_1 + m_2)^{1/5}$ is the chirp mass of the simulated binary black hole system with $m_1$ and $m_2$ being the componnent masses, $\chi_{\text{eff}}$ is the effective aligned spin, $t_0$ and $\phi_0$ denote the merger time and reference phase, and $C$ represents the parameters ($M_{\rm LV}$ and $M_{\rm PV}$) associated with Lorentz and parity violations. In our analysis, we consider eight different frequency-dependent models. For Lorentz violation, we constrain $M_{\text{LV}}$ separately for frequency-dependent damping effect with $\beta_{\bar \nu}=2$ and nonlinear dispersion relation with $\beta_{\bar \mu} = \{-2, 2, 4\}$, respectively. The signal injection parameters are listed in Table~\ref{tab:gw_parameters}. For parity violation, we constrain $M_{\rm PV}$ separately for amplitude birefringence effect with $\beta_{ \nu}=1$ and velocity birefringence with $\beta_\mu = \{-1, 1, 3\}$, respectively. This scheme allows us to efficiently map different theoretical prior choices (through $\beta$) onto constraints of the corresponding symmetry-breaking energy scales.

Note that we restrict the values of $\beta_\nu, \beta_\mu, \beta_{\bar{\nu}}, \beta_{\bar{\mu}}$ to discrete integers, as these values correspond to distinct theoretical models rather than a continuous parameter space. These specific choices are well-motivated by various modified gravity theories involving parity and Lorentz violations; see Table I in ref.~\cite{Zhu:2023rrx} for the mapping between parameter choices and specific modified theories. The frequency dependencies, characterized by odd values of $\beta_{\mu,\nu}$ for parity-violating effects and even values of $\beta_{\bar{\mu},\bar{\nu}}$ for Lorentz-violating effects, are determined by the transformation properties of the effective operators under parity transformation. Specifically, parity-violating theories introduce terms with an odd number of spatial derivatives, resulting in odd powers of the wavenumber $k$ (i.e., odd $\beta_{\mu,\nu}$). In contrast, Lorentz-violating but parity-preserving theories involve an even number of spatial derivatives, leading to even powers of $k$ (i.e., even $\beta_{\bar{\mu},\bar{\nu}}$).

\subsection{\label{subsec: 3.2}Fisher information matrix with multiband observations}

To quantify the capability of future multiband observations of GW observations to constrain parity and Lorentz symmetry violations in the gravitational sector, we perform the matched-filter analysis with the FIM approach. This approach estimates the expected statistical uncertainties of source and beyond-GR parameters from the measured waveform, given the detector's noise characteristics.

\begin{table*}
\captionsetup{justification=raggedright,singlelinecheck=false}
\caption{\label{tab:detector_config}
Basic configurations of the future ground- and space-based gravitational wave detectors considered in this work. The table lists the geometry, sensitive frequency band, arm length, and orbit/location for each instrument.}
\begin{ruledtabular}
\begin{tabular}{lcccc}
\textbf{Detector} & \textbf{Configuration} & \textbf{Frequency Band [Hz]} & \textbf{Arm Length [km]} & \textbf{Orbit/Location} \\
\hline
Cosmic Explorer (CE) & L-shaped & $5 - 5000$ & 20 / 40 & Ground (USA/Australia) \\
Einstein Telescope (ET) & Triangle & $1 - 10000$ & 10 & Ground (Europe) \\
LISA & Triangle & $10^{-4} - 0.1$ & $2.5 \times 10^6$ & Heliocentric \\
Taiji & Triangle & $10^{-4} - 0.1$ & $3 \times 10^6$ & Heliocentric \\
TianQin & Triangle & $10^{-4} - 1$ & $1.7 \times 10^5$ & Geocentric 
\end{tabular}
\end{ruledtabular}
\end{table*}

The sensitivity of a detector is characterized by its one-sided noise PSD \( S_n(f) \). To assess the distinguishability of a signal \( h(t) \) from noise and to compute parameter-estimation precision, we define the noise-weighted inner product between two signals \( a(t) \) and \( b(t) \) \cite{Cutler:1994ys}:
\begin{equation} \label{eqn:inner}
    \left( a \middle| b \right) = 4\Re \int_{f_{\mathrm{low}}}^{f_{\mathrm{high}}} \frac{\tilde{a}(f) \tilde{b}^*(f)}{S_n(f)} \,\mathrm{d}f ,
\end{equation}
where \( \tilde{a}(f) \) is the Fourier transformation of GW signal $h(t)$, and \( * \) denotes its complex conjugation. The integration bounds are set by the detector's operational band: \( f_{\text{low}} \) is typically 0.1 mHz for space-based detectors (e.g., LISA) and 1–20 Hz for ground-based instruments (e.g., CE, ET), set by seismic noise. The upper frequency limit of the integration, denoted as $f_{\text{high}}$, is determined exclusively by the high-frequency cutoff of the specific detector's sensitivity band. This threshold ensures that the analysis is strictly confined within the valid operational bandwidth of the interferometer, disregarding any signal evolution beyond the instrument's design capabilities as Tab. \ref{tab:detector_config}.

The signal-to-noise ratio (SNR), quantifying the detectability of a waveform \( \tilde{h}(f) \), follows directly from the inner product:
\begin{equation} \label{eqn:snr}
    \rho = \sqrt{ \left( \tilde{h} \middle| \tilde{h} \right) }.
\end{equation}
The precision of estimating a parameter set \( \theta \) is governed by the FIM \( \Gamma_{ij} \), which is derived from the same inner product by measuring the waveform's sensitivity to parameter changes:
\begin{equation}\label{eqn:FIM}
    \Gamma _{ij} = \left( \frac{\partial \tilde{h}}{\partial \theta_i} \middle| \frac{\partial \tilde{h}}{\partial \theta_j} \right), \quad i, j = 1, \dots, 5.
\end{equation}
In the high-SNR limit, the inverse of the FIM provides the covariance matrix \( \Sigma \) of the parameters:
\begin{equation} \label{eqn:covariance}
    \Sigma_{ij} \equiv \left( \Gamma ^{-1} \right)_{ij}.
\end{equation}
The \( 1\sigma \) statistical uncertainty for a parameter \( \theta_i \) is then given by the corresponding diagonal element:
\begin{equation}\label{eqn:uncertain}
    \sigma_{\theta_i} = \sqrt{ \Sigma_{ii} }.
\end{equation}
This formalism allows us to project the constraints on both the standard binary parameters \( (\mathcal{M}_c, \chi_{\text{eff}}, t_0, \phi_0) \) and the beyond-GR coefficients \( C \) encoding Lorentz or parity violation, as defined in Eq.~(\ref{eqn:parameters}), for the simulated detections.

Now, let us turn to consider the FIM within the multiband network. Given the statistical independence of measurements from separate detectors, we combine the constraints from instruments in the detector network.  For a parameter \(\theta_i\), the total variance from the combined network is obtained as
\begin{equation}
    \sigma_{i,\mathrm{net}}^2 = \left(\sum_{k=1}^{N_d}  \Gamma^{(k)} \right)^{-1}_{ii},
\end{equation}
where \(N_d\) denotes the number of detectors and \(\Gamma^{(k)}\) is the Fisher matrix corresponding to the \(k\)-th detector. 

\section{\label{sec: 4} Results}

\begin{table*}[htbp]
\centering
\captionsetup{justification=raggedright,singlelinecheck=false}
\caption{\label{tab:single_band_comparison}
Comparison of constraints on parity and Lorentz violation energy scales ($M_{PV}$ and $M_{LV}$) from single-band observations of GW250114 and GW231123. This table shows 90\% credibleupper bounds on $M_{\rm PV}$ for $\beta_{\mu}=-1$ and $M_{\rm LV}$ for $\beta_{\bar \mu}=-2$, while lower bounds for all other cases.
The constraints are derived from five individual detectors: CE, ET, LISA, Taiji, and TianQin.
Values are in GeV.
The upper panel displays the results for the golden event GW250114, while the lower panel displays the results for the massive event GW231123.
}
\begin{ruledtabular}
\begin{tabular}{lccccc}
\textbf{Parameter} & \textbf{CE} & \textbf{ET} & \textbf{LISA} & \textbf{Taiji} & \textbf{TianQin} \\
\hline
\multicolumn{6}{l}{\rm \textbf{Event: GW250114} } \\
$\beta_{\nu} = 1$ & $2.52 \times 10^{-20}$ & $1.39 \times 10^{-20}$ & $1.22 \times 10^{-25}$ & $4.31 \times 10^{-26}$ & $6.00 \times 10^{-26}$ \\
$\beta_{\mu} = -1$ & $3.48 \times 10^{-45}$ & $5.70 \times 10^{-45}$ & $6.67 \times 10^{-44}$ & $1.05 \times 10^{-43}$ & $2.96 \times 10^{-43}$ \\
$\beta_{\mu} = 1$ & $1.35 \times 10^{1}$ & $8.12 \times 10^{0}$ & $1.40 \times 10^{-8}$ & $3.83 \times 10^{-9}$ & $5.91 \times 10^{-8}$ \\
$\beta_{\mu} = 3$ & $2.44 \times 10^{-14}$ & $2.07 \times 10^{-14}$ & $1.13 \times 10^{-19}$ & $6.97 \times 10^{-20}$ & $8.26 \times 10^{-19}$ \\
$\beta_{\bar{\nu}} = 2$ & $6.29 \times 10^{-21}$ & $4.88 \times 10^{-21}$ & $2.03 \times 10^{-25}$ & $1.06 \times 10^{-25}$ & $4.17 \times 10^{-25}$ \\
$\beta_{\bar{\mu}} = -2$ & $1.08 \times 10^{-33}$ & $1.07 \times 10^{-33}$ & $1.43 \times 10^{-34}$ & $1.42 \times 10^{-34}$ & $2.98 \times 10^{-34}$ \\
$\beta_{\bar{\mu}} = 2$ & $1.16 \times 10^{-10}$ & $9.05 \times 10^{-11}$ & $6.40 \times 10^{-17}$ & $3.19 \times 10^{-17}$ & $4.08 \times 10^{-16}$ \\
$\beta_{\bar{\mu}} = 4$ & $3.60 \times 10^{-16}$ & $3.18 \times 10^{-16}$ & $4.83 \times 10^{-21}$ & $3.33 \times 10^{-21}$ & $3.79 \times 10^{-20}$ \\
\hline
\multicolumn{6}{l}{\rm \textbf{Event: GW231123 }} \\
$\beta_{\nu} = 1$ & $4.49 \times 10^{-20}$ & $1.29 \times 10^{-20}$ & $3.08 \times 10^{-25}$ & $1.09 \times 10^{-25}$ & $1.52 \times 10^{-25}$ \\
$\beta_{\mu} = -1$ & $1.83 \times 10^{-45}$ & $4.74 \times 10^{-45}$ & $3.14 \times 10^{-44}$ & $4.93 \times 10^{-44}$ & $1.40 \times 10^{-43}$ \\
$\beta_{\mu} = 1$ & $7.20 \times 10^{0}$ & $2.27 \times 10^{0}$ & $3.62 \times 10^{-8}$ & $9.90 \times 10^{-9}$ & $1.51 \times 10^{-7}$ \\
$\beta_{\mu} = 3$ & $8.53 \times 10^{-15}$ & $5.97 \times 10^{-15}$ & $1.60 \times 10^{-19}$ & $9.84 \times 10^{-20}$ & $1.16 \times 10^{-18}$ \\
$\beta_{\bar{\nu}} = 2$ & $4.66 \times 10^{-21}$ & $2.61 \times 10^{-21}$ & $3.30 \times 10^{-25}$ & $1.73 \times 10^{-25}$ & $6.75 \times 10^{-25}$ \\
$\beta_{\bar{\mu}} = -2$ & $7.17 \times 10^{-34}$ & $8.94 \times 10^{-34}$ & $9.43 \times 10^{-35}$ & $9.34 \times 10^{-35}$ & $1.96 \times 10^{-34}$ \\
$\beta_{\bar{\mu}} = 2$ & $4.51 \times 10^{-11}$ & $2.60 \times 10^{-11}$ & $1.05 \times 10^{-16}$ & $5.23 \times 10^{-17}$ & $6.66 \times 10^{-16}$ \\
$\beta_{\bar{\mu}} = 4$ & $1.19 \times 10^{-16}$ & $9.16 \times 10^{-17}$ & $6.32 \times 10^{-21}$ & $4.36 \times 10^{-21}$ & $4.95 \times 10^{-20}$ \\
\end{tabular}
\end{ruledtabular}
\end{table*}

\begin{figure*}[t]
    \centering
    \includegraphics[width=\textwidth]{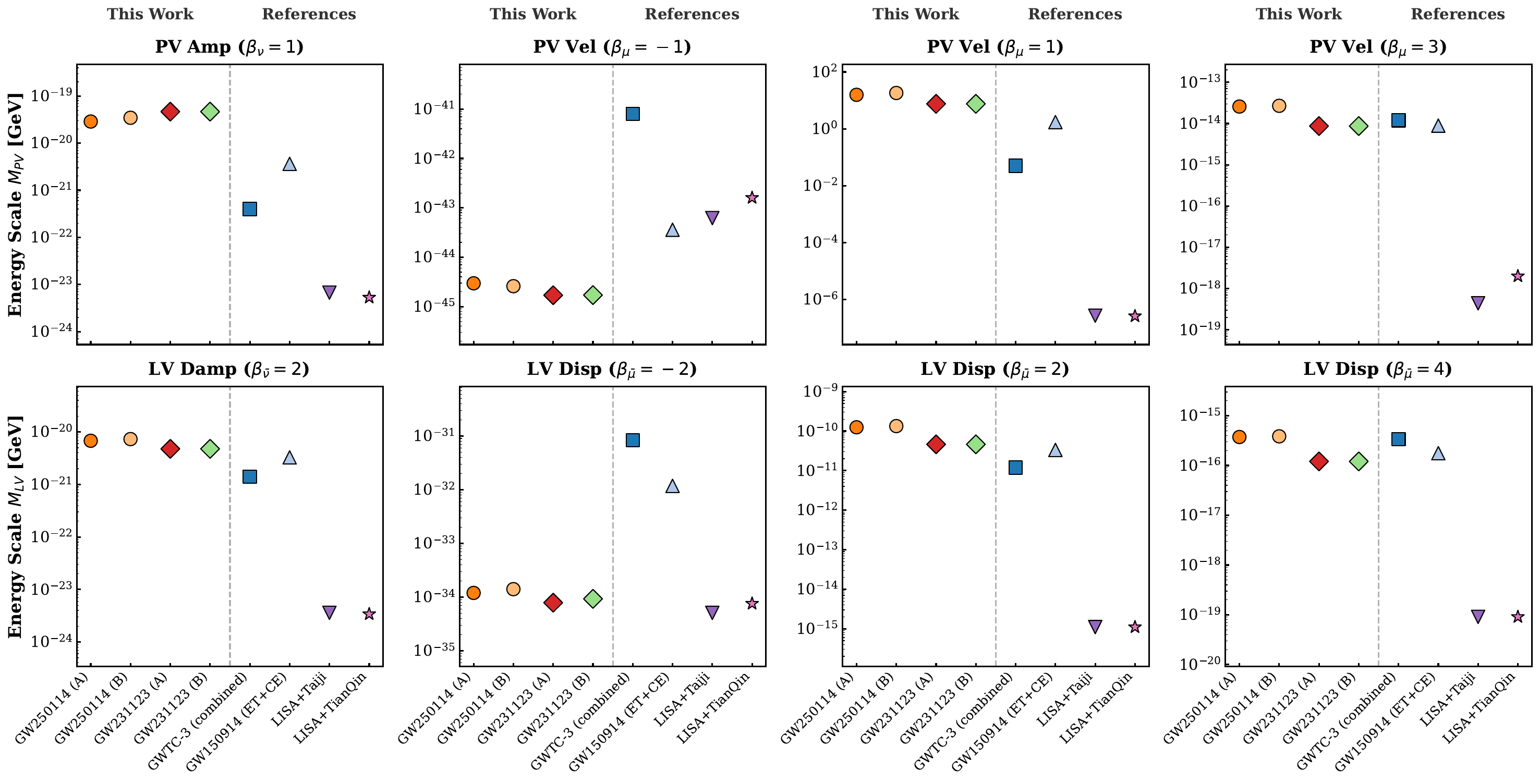}
    \captionsetup{justification=raggedright,singlelinecheck=false}
    \caption{\label{fig:comparison_with_previous_works}
 Comparison of the 90\% credible bounds on the parity-violating energy scale $M_{\rm PV}$ and Lorentz-violating energy scale $M_{\rm LV}$ derived in this work against previous benchmarks. 
 The solid markers represent the multiband constraints obtained in this work using the high-SNR event GW250114 and the massive binary GW231123. For comparison, we include the current observational limits from the LIGO-Virgo-KAGRA GWTC-3 catalog reported by ref. \cite{Zhu:2023rrx, Wang:2021gqm, Gong:2023ffb, Wang:2025fhw}, and the forecasted constraints for future ground- and space-based detectors presented by ref. \cite{Zhang:2025kcw}. The comparison demonstrates that our multiband strategy, particularly with the inclusion of the massive GW231123 event, yields constraints that significantly improve upon current baselines and are competitive with or superior to single-band future projections in the low-frequency regime. Note that in this figure, the makers represent 90\% credible upper bounds on $M_{\rm PV}$ for $\beta_{\mu}=-1$ and $M_{\rm LV}$ for $\beta_{\bar \mu}=-2$, while lower bounds for all other cases.
    }
\end{figure*}

In this section, we present constraints on parity- and Lorentz-violating energy scales from both single-detector and multiband GW observations. The analysis evaluates five detectors, the ground-based detectors CE and ET, and the space-based detectors LISA, Taiji, and TianQin; as well as two multiband observational networks: Network A (CE+ET+LISA+Taiji) and Network B (CE+ET+LISA+TianQin). To assess the impact of source properties, we employ two representative binary black hole events: the high-SNR ``golden" GW250114-like event and the massive binary GW231123-like event. The injected values of the source parameters of these two events are presented in Table~\ref{tab:gw_parameters}. In our analysis, we assume GR is the correct theory of gravity, thus the injected value for the Lorentz- and parity-violating parameter is $C=0$. The single-detector constraints are summarized in Table~\ref{tab:single_band_comparison}, the joint multiband constraints for Network A (CE+ET+LISA+Taiji) and Network B (CE+ET+LISA+TianQin) are detailed in Table~\ref{tab:multiband_comparison}, and a comparative analysis against previous studies \cite{Zhang:2025kcw, Zhu:2023rrx, Wang:2021gqm, Gong:2023ffb, Wang:2025fhw} is visualized in Fig.~\ref{fig:comparison_with_previous_works}. Note that the values represent 90\% credible upper bounds for models with negative power-law indices ($\beta_{\mu}=-1$ and $\beta_{\bar{\mu}}=-2$), and 90\% credible lower bounds for models with positive indices ($\beta_{\nu}=1, \beta_{\mu}=1, 3, \beta_{\bar{\nu}}=2, \beta_{\bar{\mu}}=2, 4$).

We first examine the constraints on Lorentz and parity violations from single-detector analysis presented in Table \ref{tab:single_band_comparison}. For the eight Lorentz- and parity-violating models we analyzed, different values of $\beta_{\bar \nu}$, $\beta_{\bar \mu}$, $\beta_\nu$, and $\beta_\mu$ correspond to different frequency-dependent amplitude and phase corrections in the waveforms. The effects with positive values of $\beta_{\bar \nu}$, $\beta_{\bar \mu}$, $\beta_\nu$, and $\beta_\mu$ are more relevant at high frequency, while the negative values are more sensitive to low frequency of GWs. From the results in Table \ref{tab:single_band_comparison}, the constraints reveal a clear frequency-dependent pattern dictated by the power-law index $\beta_{\bar \nu}$, $\beta_{\bar \mu}$, $\beta_\nu$, and $\beta_\mu$. For high-frequency modifications (positive values of $\beta_{\bar \nu}$, $\beta_{\bar \mu}$, $\beta_\nu$, and $\beta_\mu$), the ground-based detectors strongly outperform space-based ones. The constraints on $M_{\rm LV}$ and $M_{\rm PV}$ for $\beta_{\bar \nu}=2$, $\beta_{\bar \mu}=2, 4$, $\beta_\nu=1$, and $\beta_{\mu} = 1, 3$ from ET and CT are roughly stronger than those from the space-based detector by 3 -7 orders of magnitude. For the results from CE and ET, the GW250114-like event gives roughly better constraints than the GW231123-like event for most cases. This is because the GW250114-like event is lighter than the GW231123-like event, and the amplitude and phase corrections for the positive values of $\beta_{\bar \nu}$, $\beta_{\bar \mu}$, $\beta_\nu$, and $\beta_\mu$ are larger at high frequencies.

In contrast, for low-frequency effect (for example, Lorentz-violating dispersion with $\beta_{\bar{\mu}}=-2$), space-based detectors provide superior constraints, with sensitivity scaling with both source mass and detector arm length. The massive GW231123-like event yields tighter bound than GW250114-like event across all space-based instruments; for example, LISA constrains $M_{\mathrm{LV}} < 9.43 \times 10^{-35}$ GeV for GW231123, compared to $1.43 \times 10^{-34}$ GeV for GW250114. Longer-baseline space detectors generally give stronger limits: Taiji ($\sim 3\times10^6$ km) and LISA ($\sim 2.5\times10^6$ km) outperform TianQin ($\sim 1.7\times10^5$ km) for the same event and modification term.

Building upon these single-detector results, we then evaluate the constraints obtained from analysis with the joint multiband configurations summarized in Table~\ref{tab:multiband_comparison} and presented in Fig.~\ref{fig:comparison_with_previous_works}. Multiband networks combine the complementary sensitivity windows of space- and ground-based observatories, typically yielding constraints that surpass any single instrument. For both the two multiband networks, the network A (ET+CE+LISA+Taiji) and network B (ET+CE+LISA+TianQin), we observe that multiband observations from the two networks improve the results achieved by the single detectors, providing stronger constraints on Lorentz and parity violations across all eight models we analyzed.

For comparison with earlier studies, we also present in Fig.~\ref{fig:comparison_with_previous_works} the constraints reported in refs.~\cite{Zhang:2025kcw, Zhu:2023rrx, Wang:2021gqm, Gong:2023ffb, Wang:2025fhw}. Those analyses derived constraints on the same eight Lorentz- and parity-violating models considered here by combining multiple binary coalescence events from GWTC-3. Concretely, the case $\beta_\mu = 1$ was studied in ref.~\cite{Wang:2021gqm}, $\beta_{\bar{\mu}} = -2$ in ref.~\cite{Gong:2023ffb, Wang:2025fhw}, and $\beta_{\nu}=1$, $\beta_{\mu}=-1, 3$, $\beta_{\bar{\nu}}=2$, $\beta_{\bar{\mu}}=2, 4$ in ref.~\cite{Zhu:2023rrx}.

As illustrated in Fig.~\ref{fig:comparison_with_previous_works}, multiband observations using the two detector networks can improve these existing bounds by 1–3 orders of magnitude. When compared specifically to the results of ref.~\cite{Zhang:2025kcw}, which considered ET, CE, LISA, Taiji, and Tianqin for different types of sources, our analysis yields moderately stronger constraints for some models while remaining comparable for others. For instance, on the low-frequency-sensitive model $\beta_{\bar{\mu}}=-2$, constraints derived from the massive event GW231123 improve upon previous limits by nearly an order of magnitude. Conversely, for high-frequency sensitive models such as $\beta_{\mu}=1$, bounds obtained from the golden event GW250114 align with established ground-based benchmarks.

These comparisons confirm that carefully selected individual binary systems, when observed simultaneously across multiple frequency bands, can deliver exceptionally stringent tests of fundamental symmetries, in certain regimes even surpassing statistical constraints obtained from large population studies. Future multiband networks will therefore provide a powerful and complementary pathway for probing departures from GR across a wide range of energy scales.

It should be noted that the energy scales $M_{\rm PV}$ and $M_{\rm LV}$ are mathematically degenerate with the dimensionless coefficients $\alpha$ that characterize the coupling strength of specific modified gravity theories. In our parameterized treatment, these dimensionless coefficients are effectively absorbed into the definitions of $M_{\rm PV}$ and $M_{\rm LV}$ by assuming $\alpha$ to be a constant of order unity across the relevant frequency bands. Consequently, the bounds derived in this work on these energy scales can be directly mapped to the specific parameters of various theoretical models by unfolding this degeneracy. This ensures that our results provide a versatile and robust test for a wide range of gravitational extensions of GR.

\begin{table*}[htbp]
\centering
\captionsetup{justification=raggedright,singlelinecheck=false}
\caption{\label{tab:multiband_comparison}
Comparison of constraints on parity and Lorentz violation energy scales from joint multiband observations. This table shows 90\% credible upper bounds on $M_{\rm PV}$ for $\beta_{\mu}=-1$ and $M_{\rm LV}$ for $\beta_{\bar \mu}=-2$, while lower bounds for all other cases. The table displays results for two detector networks: \textbf{Network A} (CE + ET + LISA + Taiji) and \textbf{Network B} (CE + ET + LISA + TianQin). Constraints are provided for both the golden event GW250114 and the massive event GW231123. Values are in GeV.
}
\begin{ruledtabular}
\begin{tabular}{l c c c c}
\textbf{Parameter} & \multicolumn{2}{c}{\textbf{GW250114}} & \multicolumn{2}{c}{\textbf{GW231123}} \\
\cline{2-3} \cline{4-5}
 & \textbf{Network A} & \textbf{Network B} & \textbf{Network A} & \textbf{Network B} \\
\hline
$\beta_{\nu} = 1$ & $2.88 \times 10^{-20}$ & $3.46 \times 10^{-20}$ & $4.67 \times 10^{-20}$ & $4.67 \times 10^{-20}$ \\
$\beta_{\mu} = -1$ & $2.96 \times 10^{-45}$ & $2.59 \times 10^{-45}$ & $1.70 \times 10^{-45}$ & $1.71 \times 10^{-45}$ \\
$\beta_{\mu} = 1$ & $1.57 \times 10^{1}$ & $1.82 \times 10^{1}$ & $7.55 \times 10^{0}$ & $7.55 \times 10^{0}$ \\
$\beta_{\mu} = 3$ & $2.58 \times 10^{-14}$ & $2.69 \times 10^{-14}$ & $8.69 \times 10^{-15}$ & $8.69 \times 10^{-15}$ \\
\hline
$\beta_{\bar{\nu}} = 2$ & $6.80 \times 10^{-21}$ & $7.32 \times 10^{-21}$ & $4.77 \times 10^{-21}$ & $4.77 \times 10^{-21}$ \\
$\beta_{\bar{\mu}} = -2$ & $1.20 \times 10^{-34}$ & $1.41 \times 10^{-34}$ & $7.89 \times 10^{-35}$ & $9.31 \times 10^{-35}$ \\
$\beta_{\bar{\mu}} = 2$ & $1.25 \times 10^{-10}$ & $1.34 \times 10^{-10}$ & $4.63 \times 10^{-11}$ & $4.63 \times 10^{-11}$ \\
$\beta_{\bar{\mu}} = 4$ & $3.74 \times 10^{-16}$ & $3.86 \times 10^{-16}$ & $1.21 \times 10^{-16}$ & $1.21 \times 10^{-16}$ \\
\end{tabular}
\end{ruledtabular}
\end{table*}

\section{\label{sec: 5}Conclusion and Outlook}

In this work, we have performed a comprehensive investigation into the capability of future ground- and space-based GW detector networks to constrain violations of fundamental symmetries in gravity. Through FIM analysis of modified waveforms with parity-violating and Lorentz-violating effects, we demonstrated the transformative capability of multiband observations by analyzing two representative binary black hole sources: the high-SNR ``golden event" GW250114-like event and the massive binary GW231123-like event. The core of this strategy lies in exploiting the complementary frequency coverage of different detectors to track a single source across its entire evolution. Stellar-mass binary black holes such as GW250114 ($M_{\rm tot} \approx 65 M_\odot$) inspiral for years in the millihertz band observable by space-based detectors (LISA, Taiji, TianQin) before entering the high-frequency band of ground-based observatories (ET, CE).

Our single-detector analysis reveals a clear frequency-dependent pattern. For high-frequency modifications (\(\beta > 0\)), ground-based detectors (CE, ET) provide constraints 3–7 orders of magnitude stronger than space-based ones. The lighter, louder GW250114 gives better bounds in this regime. For low-frequency modifications (\(\beta < 0\)), space-based detectors (LISA, Taiji, TianQin) dominate. The massive GW231123 yields tighter constraints, and longer-baseline detectors (Taiji, LISA) outperform shorter ones (TianQin).

We evaluated two multiband networks: Network A (CE+ET+LISA+Taiji) and Network B (CE+ET+LISA+TianQin). Both networks improve upon current bounds by several orders of magnitude. Compared to previous studies using multiple events in GWTC-3 detected by LIGO, Virgo, and KAGRA detectors, our multiband analysis with carefully selected individual sources provides competitive or stronger constraints. In particular, for the low-frequency model \(\beta_{\bar{\mu}} = -2\), GW231123 improves previous limits by nearly an order of magnitude.

Looking ahead, the strategic selection of sources GW250114-like events, whose evolution naturally bridges the millihertz and kilohertz bands, will be essential for realizing the full potential of multiband GW observations. The concurrent operation of next-generation space- and ground-based observatories will not only sharpen limits on parity and Lorentz violation but also open new avenues for probing strong-field gravity. Future extensions of this framework could include eccentric orbits, higher-order waveform modes, and population-level multiband analyses, further establishing GW observations as a precision tool for fundamental physics.

\begin{acknowledgments}

This work is supported by the National Key Research and Development Program of China under Grant No. 2020YFC2201503, the National Natural Science Foundation of China under Grants No.~12275238 and No.~11675143, the Zhejiang Provincial Natural Science Foundation of China under Grants No.~LR21A050001 and No.~LY20A050002, and the Fundamental Research Funds for the Provincial Universities of Zhejiang in China under Grant No.~RF-A2019015. In addition, we used Google Gemini (https://gemini.google.com/) and DeepSeek (https://chat.deepseek.com/) for language polishing and proofreading of this manuscript.
 
The data analyses and results visualization in this work made use of   \texttt{LALSuite} \cite{LALSuite, swiglal}, \texttt{Numpy} \cite{harris2020array}, \texttt{Scipy} \cite{2020SciPy-NMeth}, \texttt{PyCBC} library \cite{PyCBC}, and \texttt{matplotlib} \cite{Hunter:2007}.

\end{acknowledgments}



\bibliography{apssamp}

\end{document}